# Decoupling control parameter method to study the coupling characteristics of subsystems in the feed system

Dongsheng Zhang[1]，Xuesong Wang[1]*, Tingting Zhang[2]

[1]Department of mechanical engineering, Xi'an Jiaotong University, Xi'an 710064 China

[2]AVIC The First Aircraft Institute, Xi'an 710089 China

Corresponding author: Xuesong Wang(3121301048@stu.xjtu.edu.cn)

Abstract

When developing high-speed and high-precision CNC machine tools, subsystem coupling effects must be considered while designing the feed system to maximize its dynamic performance. Currently, the influence of changes in control parameters on the matching characteristics of each subsystem was not yet considered when studying the coupling relationship between subsystems. Therefore, it is difficult to define the law of action between subsystems under the interference of control parameters. Hence, a new method was proposed in this paper aiming to isolate the disturbance of control parameters and to highlight the actions between electromechanical subsystems. This was achieved by optimizing the servo control parameters for each configuration of electromechanical parameters, ensuring that the feed system performance is optimal for the observed configuration. This approach eliminated the influence of the control subsystem on the electromechanical coupling relationship. The approach effectiveness was verified via the integrated model of the feed system. As such, this paper provides a reliable method to further study the coupling mechanism of subsystems, revealing the mechanism behind the dynamic design of feed systems.

## 1 Introduction

High speed and high precision were always the aim when developing CNC machine tools(1) and feed systems. Further, as CNC machine tool actuator, the feed system largely determines the resulting machining quality. Therefore, when designing the feed system, it is necessary to ensure good dynamic performance. The available design methods hardly meet the demand, with problems related to unquantified and unclear underlying mechanisms persisting(2). To study the dynamic optimization design of the feed system in high-speed and high-precision CNC machine tools, it is first needed to study the coupling mechanism between subsystems.

A lot of research was done on the coupling mechanism between the feeding system components(3,4). Szolc et al. studied the electromechanical coupling between the open-loop drive system of the induction motor and the rotating load. It was found that, under the harmonic excitation of load torque, the electromechanical system natural frequency changes relative to the natural frequency of the mechanical system. This is due to the variation in the electromagnetic field between the drive motor rotor and stator (5). Lu et al. analyzed the effect of high speeds and accelerations on the electromechanical coupling of the feed system. The results have shown that, under high speed and high acceleration, the time-varying characteristics of the mechanical system mass distribution and stiffness are more significant. Furthermore, the harmonic components of the servo output force increase, along with the amplitude. Additionally, in that case, the cutting force also has a more complex frequency component, resulting in a more complex electromechanical coupling (6). Yang et al. analyzed the coupling effect between the motor and the motion process, finding that the system displacement fluctuations caused by the motor thrust harmonics will gradually worsen with the increase in feed speed and acceleration. Hence, when designing a high-speed, high-precision feed system, the motion parameters should be selected in a way to avoid resonant speeds which yield poor accuracy(7–11).

The above-reviewed content focused on the matching characteristics between the mechanical structure, the drive motor, and the motion process. The results have shown that the coupling effect of each subsystem significantly affects the system performance. However, the studies have not considered if the change of control parameters affects the influence of electromechanical coupling characteristics on the system performance. By analyzing the closed-loop control schematic of the feed system, (see Figure 1-1), the specific transfer function was obtained and shown in Equation(1-1). In equation, $R(s)$ represents the motion process, $G_c(s)$ is the servo control, $G_a(s)$ represents the drive motor, and $G_p(s)$ is a mechanical structure. The control parameters $G_c(s)$ have the same effect on the output $Y(s)$ as the motor parameters and mechanical structure parameters.

$$Y(s) = \frac{G_c(s) \cdot G_a(s) \cdot G_p(s)}{1 + G_c(s) \cdot G_a(s) \cdot G_p(s)} R(s) \tag{1-1}$$

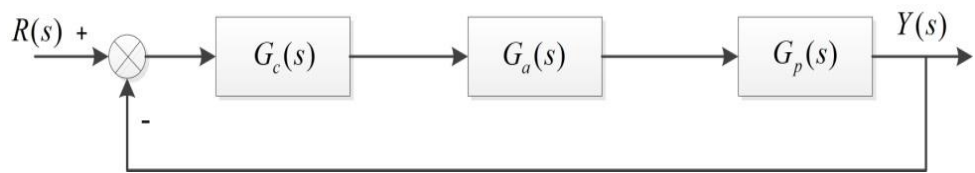

Figure 1-1 Closed-loop control schematic diagram of the feed system

The influence of control parameters on mechanical structure and driving motor characteristics was also studied. Kim et al. analyzed the effect of mechanical design variables (ball screw diameter and lead) and control parameters on the performance of servo systems. The system performance was affected by the parameter characteristics of both mechanical and control components(12). Liu et al. analyzed the effect of control parameters on the mechanical and servo systems from a frequency domain perspective. The higher-order frequencies of certain mechanical structures were excited as the speed gain was increased. The servo thrust varied with the feed speed, while the servo thrust spectrum was coupled with the inherent frequencies of the mechanical structure at a specific feed system speed, decreasing the performance(13). Hence, the change in control parameters will affect the coupling relationship between the mechanical system and the drive motor; changing the control parameters might result in opposite electromechanical coupling action laws.

Therefore, to clarify the laws governing the feed system dynamic performance with the electromechanical coupling characteristics, it is first necessary to clarify the underlying mechanism behind the feed system dynamics. It is critical to eliminate the interference of control parameters on the research results. In this paper, the author proposed a new approach to eliminate the influence of control parameters on the electromechanical coupling characteristics. The approach is based on the optimization of the servo parameters, which are optimized for each electromechanical parameter configuration to achieve the current optimal performance of the feed system. Next, through a comparison of the optimal performance variations under various parameter configurations, the coupling characteristics of each subsystem can be explored. The approach itself provides a reliable method to further study the coupling mechanism of subsystems, revealing the underlying mechanism behind the dynamic design of feed systems.

The concept of integrated modeling for the electromechanical control and motion process of feed systems was introduced in Chapter 2. Chapter 3 covers the method for evaluating the motion performance of feed systems. In Chapter 4, the influence of control parameters on electromechanical coupling characteristics was discussed, while a decoupling method for control parameters was shown in Chapter 5. Finally, Chapter 6 concludes the paper with a discussion and final remarks.

# 2  Integrated modeling of feed system electromechanical control and motion process

To carry out the study at hand, it is first necessary to establish an integrated model consisting of a mechanical structure, drive motor, servo control, and motion process. The single-axis feed system was used as a research object; Solidwork was used to create the mechanical model, which was then transformed into a visual dynamics model through MATLAB/Simscape. Next, MATLAB was used to build the servo control system and the motion process system, as well as to finalize the joint modeling of the mechanical system, control system, and motion process. All of the above was a base to ensure that the model is as close as possible to the actual working conditions of the machine tool(14).

## 2.1 Modeling of mechanical transmission systems

In this paper, the mechanical structure was modeled via Solidwork. Due to the complexity of the actual servo feed system mechanical model, the model was simplified to reduce computational costs. This was done while ensuring the kinematics and dynamics characteristics of the feed system. The resulting mechanical transmission system model is shown in Figure 2-1.

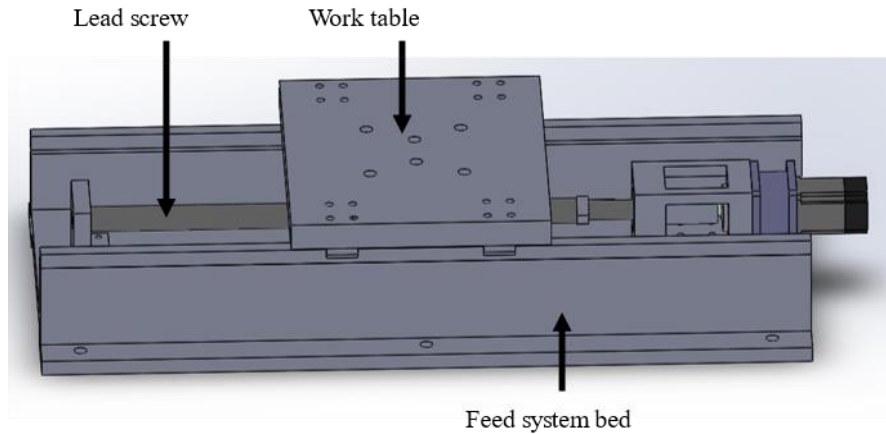

Figure 2-1 Three-dimensional feed system model

The constraint relationships for each component are shown in Table 2-1 Constraint relations of mechanical components; static parts were not listed.

Table 2-1 Constraint relations of mechanical components

| Component 1 | Component 2 | Constraint relationships | Constrained degrees of freedom |
| --- | --- | --- | --- |
| Motor stator | Motor rotor | Rotation pair | 5 |
| Lead screw | Screw nut | Screw pair | 4 |
| Bearing | Lead screw | Cylindrical pair | 4 |
| Guide rail | Slider | Sliding pair | 5 |

The 3D Solidwork model was output as an XML format file and imported into MATLAB/Simscape environment to carry out the co-simulation analysis with the servo control system. The model imported into MATLAB/Simscape retains the constraint relationship of the original parts. The results are shown in Figure 2-2.

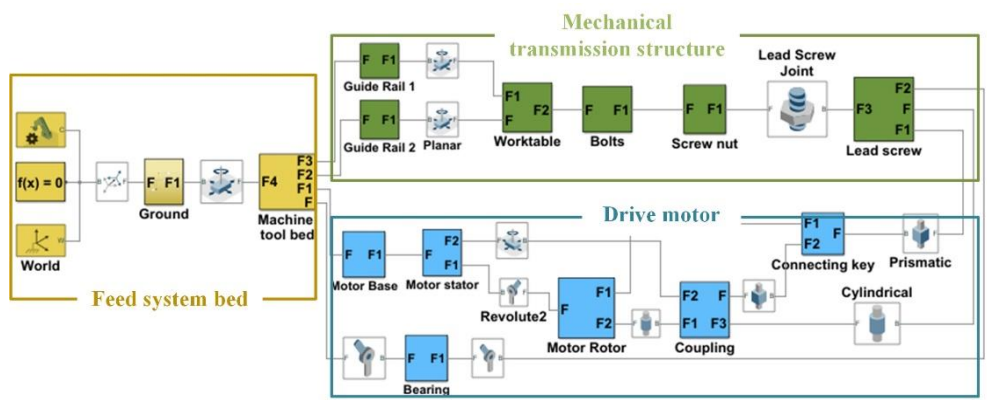

Figure 2-2 Virtual prototype mechanical model

In the above-mentioned figure, among these kinematic pairs, Revolute pair represents a single-degree-of-freedom rotating pair, Prismatic is a single-degree-of-freedom sliding pair, and Planar is a 2-degree-of-freedom plane sliding pair. These relation pairs were used to connect rigid bodies and limit their degrees of freedom. In the Simscape environment, the virtual feed system mechanical model was built.

The mechanical model of the feed system transmission includes the worktable, lead screw, and motor, among other components. The main mechanical characteristic parameters refer to the specification parameters of the servo mechanism platform(15), as shown in Table 2-2.

Table 2-2 Characteristic parameters of mechanical structures

| Variables | Meaning | Values | Unit |
|---|---|---|---|
| $K$ | Lead screw stiffness | 612 | $N \cdot m \cdot rad^{-1}$ |
| $J_L$ | Load inertia | 45.5 | $kg \cdot cm^2$ |
| $B$ | Damping | 0.0288 | $kg \cdot m^2 \cdot s^{-1}$ |
| $R$ | Lead screw transmission coefficient | $10/2\pi$ | $mm \cdot rad^{-1}$ |

To study the influence of control parameters on the electromechanical coupling characteristics, the load inertia was taken as constant $1 \leq r \leq (2.5 \sim 5)$, in accordance with the empirical inertia ratios used in machine tool design(16). Six servo motors were selected to obtain different configurations of electromechanical parameters; the specific characteristics of motors are shown in Table 2-3.

Table 2-3 Servo motor characteristics

| Serial number | Motor Model | Maximum Torque T/N m | Rotor inertia $J_m$/kg cm$^2$ | Inertia Ratio |
|---|---|---|---|---|
| 1 | ISMH3-44C15CD | 71.1 | 88.9 | 0.5 |
| 2 | ISMH3-29C15CD | 37.2 | 55 | 0.8 |
| 3 | ISMH3-18C15CD | 28.75 | 25.5 | 1.8 |
| 4 | ISMH3-13C15CD | 20.85 | 19.3 | 2.4 |
| 5 | ISMH3-85B15CD | 13.5 | 13 | 3.5 |
| 6 | 1MH3-50B15CB | 9.6 | 11.01 | 4.1 |

2.2 Modeling servo control systems

Servo controllers usually use multi-loop PID control consisting of position, velocity, and current loops. The current loops are used to enhance the anti-interference capability and dynamic tracking performance of the system. A two-loop PID was used in this study; the velocity feedforward control was built in MATLAB (see Figure 2-3), consisting of speed loop and position loop structures from the inside to the outside.

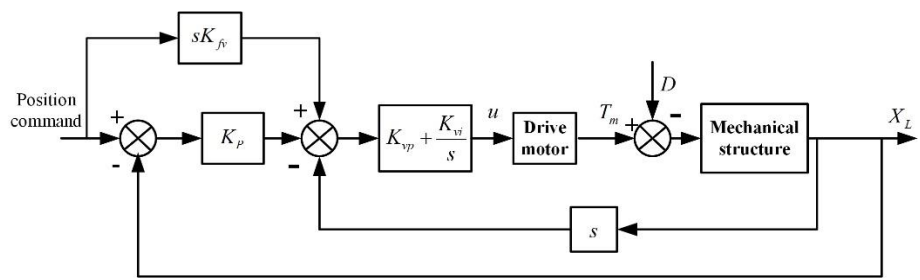

Figure 2-3 Controller block diagram

In the above-mentioned figure, $K_{fv}$ is the speed feedforward proportional coefficient, $K_p$ is the position loop proportional coefficient, $K_{vp}$ is the speed loop proportional coefficient, $K_{vi}$ is the speed loop integral coefficient, D is the load torque, and $T_m$ is the motor output torque. The position loop uses proportional control to achieve accurate positioning control and to ensure the dynamic tracking performance of the system. The speed loop uses a PI controller to enhance the anti-disturbance capability of the system and suppress speed fluctuations.

2.3 Modeling of motion processes

In this study, performance indicators such as maximum speed/maximum acceleration, commonly used in machine tools, were used in combination with a trapezoidal speed planning curve, as shown in Figure 2-4. The curve was divided into three phases: acceleration, uniform speed, and deceleration.

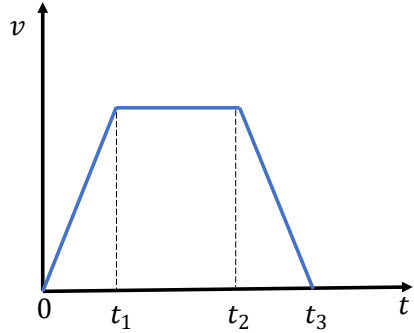

Figure 2-4 Trapezoidal speed planning curve

The specific calculation processes are as follows:
(1) Acceleration($0 \sim t_1$):

$$v(t) = at, \quad s_a = \frac{1}{2}at_1^2 \tag{2-1}$$

(2) Uniform movement($t_1 \sim t_2$):

$$v(t) = v, \quad s_u = v(t_2 - t_1) \tag{2-2}$$

(3) Deceleration($t_2 \sim t_3$):

$$v(t) = v - a(t_3 - t_2), \quad s_d = \frac{1}{2}a(t_3 - t_2)^2 \tag{2-3}$$

where $a$ is the acceleration, $v$ is the uniform velocity, $s_a$ is the acceleration distance, $s_u$ is the distance covered while traveling at a constant velocity, and $s_d$ is the deceleration distance.

The motion parameters of the feed system are determined based on the actual machine requirements, as shown in Table 2-4.

Table 2-4 Motion process parameters

| Parameters | Displacement | Speed | Acceleration | $t_1$ | $t_2$ | $t_3$ |
|---|---|---|---|---|---|---|
| Values | 200 mm | 100 mm/s | 5 m/s² | 0.02 | 2 | 2.02 |

In this paper, the S-function module of MATLAB was used to write the interpolation algorithm based on the trapezoidal velocity planning curve in C. Therefore, the table can complete a reciprocating motion with a distance of 200 mm at a speed of 100 mm/s, and an acceleration of 5 m/s², achieving smooth operation of the feed system.

2.4 Integrated model of the feed system

The mechanical transmission system, servo control system, and motion process modeling of the feed system were made using Solidwork and MATLAB/Simscape, respectively. The integrated feed system model was built through the MATLAB communication interface (17).

In the feed system, the servo motor is the power unit driving the table motion. It converts displacement command into the torque needed to drive the table. Therefore, in the virtual environment, the servo motor is represented as the interface for data exchange between the control and the mechanical systems. The control system inputs the torque into the mechanical structure, while the mechanical structure outputs the speed and table displacement. Both values are fed back into the control system, forming a closed loop to achieve the joint simulation of the feed system.

After defining the input and output variables of the mechanical and the control systems, a joint simulation model of the feed system was built in MATLAB/Simscape, as shown in Figure 2-5.

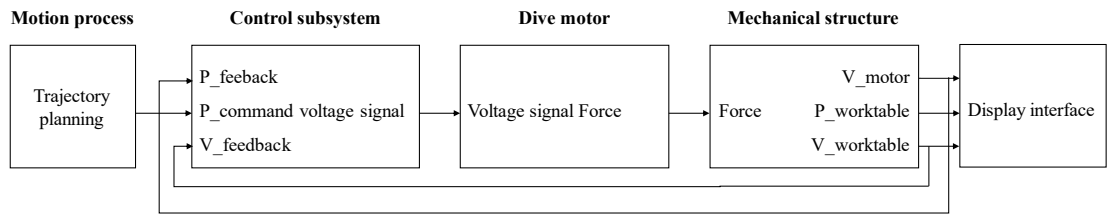

Figure 2-5 Experimental model of the feed system

Next, to evaluate the influence of the electromechanical coupling characteristics on the feed system motion performance, it is necessary to define an evaluation method. The method must be able to systematically and comprehensively describe the strengths and weaknesses of the CNC machine tools concerning key performance indicators.

# 3 Performance evaluation method of motion performance of feed system

In engineering applications, positioning accuracy is generally used to evaluate the accuracy of the CNC machine tool during motion. It is a static evaluation method unable to evaluate the transient accuracy of the feed system during acceleration and deceleration(18). Therefore, in this paper, the degree of conformity between the motion command and the actual output was used as the evaluation criterion. The position error, velocity error, and velocity fluctuation were used to rate the dynamic characteristics of the feed system. Weight factors were assigned to each performance index as the comprehensive evaluation index of the system. The characterization method is shown in Equation (3-1):

$$W = 0.5*\max(Err\_p) + 0.25*\max(Err\_v) + 0.25*Vars\_v \qquad (3\text{-}1)$$

The evaluation indexes $W$ consider position and speed accuracy; the former is assessed by the maximum value of position error $max(Err\_p)$, while the latter mainly considers the maximum value of speed error $max(Err\_v)$ and speed fluctuation $Vars\_v$. Both values can characterize the motion accuracy of the feed system in the steady state and transient process. Finally, the smaller the value $W$, the higher the feed system motion accuracy.

In this case, the velocity fluctuation $Vars\_v$ was calculated using segmentation. The velocity error is vastly different between the acceleration and deceleration phases and the uniform velocity phase; therefore, the segmentation is calculated according to the velocity process. The standard deviation of the velocity error within each segment was first calculated and then summed. The summation was processed by multiplying the velocity fluctuation with the proportion of the period in the whole cycle. In other words, the average velocity fluctuation per unit of time was obtained, as follows:

$$Vars\_v = \sum_i std_i(Err\_v) * t_i / T \qquad (3\text{-}2)$$

where $std_i(Err\_v)$ is the standard deviation of the speed of the *i*-th segment, $t_i$ is the running time of the *i*-th segment, and $T$ represents the time of the whole motion cycle.

Considering all the system performance indicators, the proposed expression was used as a criterion to analyze the degree of influence of the control parameters on the law of electromechanical coupling.

# 4 Influence of control parameters on electromechanical coupling characteristics

Based on the integrated feed system model developed in Chapter 2, the drive motors were varied to obtain different electromechanical configurations, constituting different load inertia ratios. The worktable was made to complete the reciprocating motion of 200 mm with a speed of 100 mm/s and an acceleration of 5 m/s². The motion performance of the feed system was analyzed through this process.

The operations were carried out as follows: the load inertia of the feed system remains constant, while the motor parameters are changed according to Table 2-3. When the motor parameters are changed, the load inertia ratio of the system changes accordingly. The influence of the inertial ratio on the system performance was taken as the evaluation standard for the coupling characteristics of electromechanical subsystems. The aim was to assess whether the change in control parameters will affect the way the electromechanical coupling characteristics influence the system performance, which was done by changing the control parameters as shown in Table 4-1. This allowed the authors to analyze the changes in the feed system performance under these three groups of control parameters, each corresponding to different inertia ratios, i.e., the effect of control parameters on the law of electromechanical coupling.

Table 4-1 Settings of servo control parameters

| Control parameters<br>Serial number | $K_p$ | $K_{vp}$ | $K_{vi}$ | $K_{fv}$ |
|---|---|---|---|---|
| 1 | 10 | 20 | 50 | 1 |
| 2 | 50 | 20 | 5 | 0.5 |
| 3 | 200 | 20 | 5 | 1 |

The results are shown in Figure 4-1:

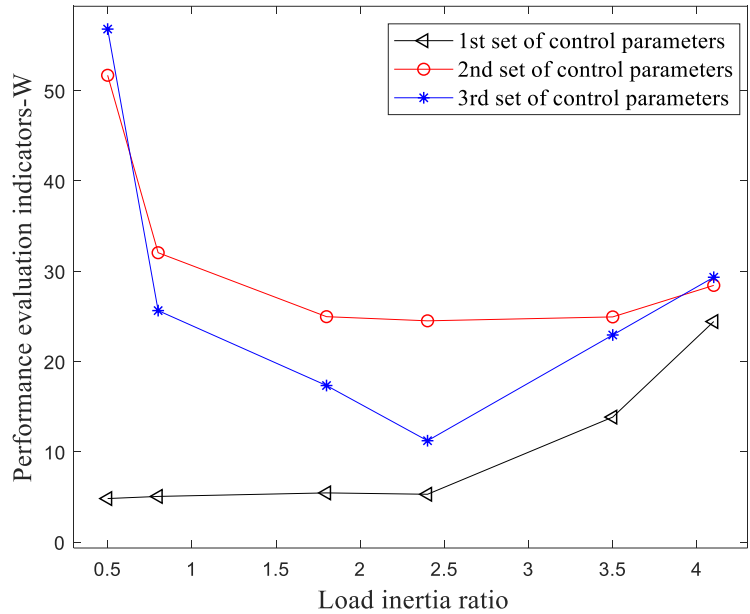

Figure 4-1 Effect of inertia ratio on system performance with different control parameters

The figure above illustrates the variations between the performances of different motors under various control parameters. The horizontal coordinate represents the load inertia ratio, while the vertical coordinate is the system comprehensive evaluation index $W$. The symbols ◁ , ○ , and ✻ lines in the figure correspond to group 1, group 2, and group 3 control parameters, respectively. By analyzing the change of "load inertia ratio - system evaluation index W" for the three groups of control parameters. For the first group, the system performance gradually improves and finally stabilizes as the inertia ratio decreases. In the second group, the system performance gradually improves and finally stabilizes as the inertia ratio increases. Finally, for the third group performance, the system performance gradually worsened as the inertia ratio became either too large or too small.

When the control parameters change, corresponding inertia ratios alter the system performance. If the specific value of the system performance and the overall trend have changed, the results imply that the control parameters affect the electromechanical coupling characteristics, and subsequently, the system performance law. Therefore, to study the coupling characteristics between the motor and the mechanical structure, it is firstly necessary to eliminate the influence of the control parameters on the electromechanical coupling characteristics. For this reason, this paper further studied the ways to eliminate the influence of the control parameters on the study results.

## 5  Decoupling method for control parameters

It was proposed that the control parameters can be optimized for each group of feed system electromechanical configurations so the system can reach the corresponding optimal state. This was done to isolate the influence of control parameters on the system performance and highlight the coupling law between electromechanical subsystems, as shown in Figure 5-1.

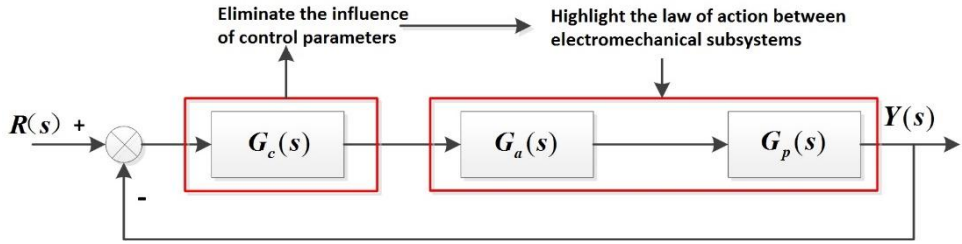

Figure 5-1 Schematic diagram of the control parameter decoupling method

The control module $G_c(s)$ includes several servo control parameters, such as velocity feedforward proportionality coefficient $K_{fv}$, position loop proportionality coefficient $K_p$, velocity loop proportionality coefficient $K_{vp}$, and velocity loop integration coefficient $K_{vi}$. On the other hand, the evaluation index W contains system performance indicators (position error, velocity error, velocity fluctuation). Therefore, the servo parameter optimization is a multi-objective, multi-parameter optimization problem.

To ensure the credibility of the servo parameter optimization results, the stability of the optimization algorithm itself must be guaranteed. Therefore, a population intelligence optimization algorithm titled the fireworks algorithm was used. The fireworks algorithm is currently used in multi-objective optimization problems such as software project scheduling and vehicle path planning, primarily due to its global and local search characteristics(19,20). Next, the servo-parameter optimization capability of the fireworks algorithm will be evaluated and validated as the decoupling method for the control parameters.

5.1 Evaluation of fireworks algorithm optimization capability

(1) The fireworks algorithm is an optimization algorithm proposed in 2010 which iterates through initial N fireworks. It comprises the explosion operator, the variation operator, the mapping rule, and the selection strategy. The mapping rule is used to ensure that the fireworks are in the feasible domain, while the elite random selection strategy is to achieve the selection of the offspring fireworks until the termination condition is met. The fireworks algorithm execution flowchart is shown in Figure 5-2.

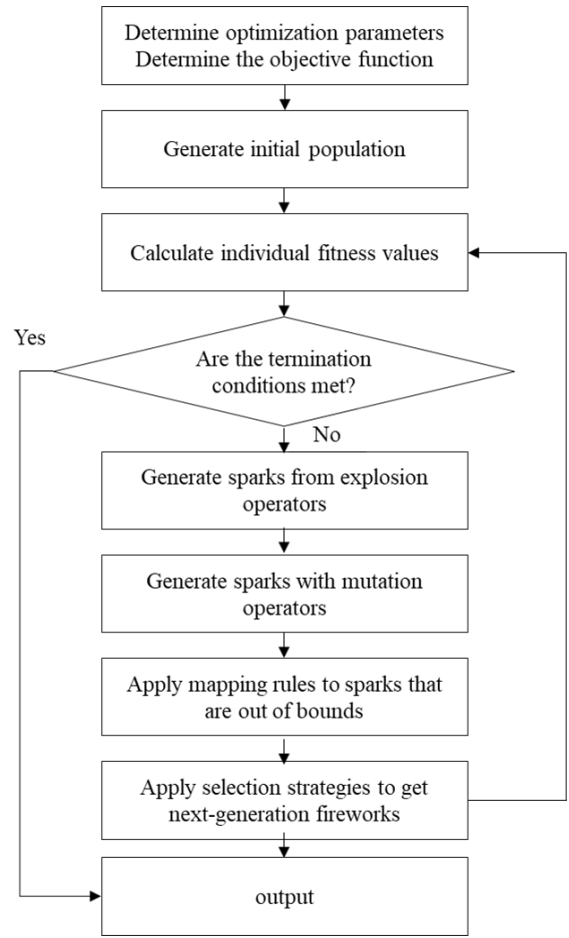

Figure 5-2 Flowchart of the fireworks algorithm

In each iteration of the fireworks algorithm, each firework explodes at different coordinate ranges, meaning that the search is carried out at various coordinate ranges. The distributed parallelism and multiplicity ensure that the algorithm jumps out of the local extreme point and converges towards the global optimal solution.

(2) The worktable with the maximum speed of $100 \, \text{mm/s}$, the acceleration of $5 \, \text{m/s}^2$ motion process, completes the distance of $200 \, \text{mm}$ reciprocating motion. The control parameters $K_p$, $K_{vi}$, $K_{vp}$, and $K_{fv}$ were optimized using the fireworks algorithm and the control parameters are given in Table 5-1.

With the constant system parameters, the control parameters were optimized six times using the fireworks algorithm and the traditional genetic algorithm. The algorithm parameters are given in Table 5-2 and Table 5-3. The changes in system performance corresponding to the six optimization results were observed, as shown in Figure 5-3. The results show that the fireworks algorithm has good stability, yielding good quality optimization results. It is a very practical multi-objective, multi-parameter rectification method

Table 5-1 Control parameter Settings

| Parameters | Value |
|---|---|
| $K_p$ | [0,200] |

| | | |
|---|---|---|
| | $K_{vi}$ | [0,20] |
| | $K_{vp}$ | [0,5] |
| | $K_{fv}$ | [0.5,1] |

Table 5-2 Fireworks algorithm parameter settings

| Parameters | Generations | Spark number | Mutaion spark number | Explosion number | Explosion radius |
|---|---|---|---|---|---|
| Value | 50 | 20 | 5 | 6 | 5 |

Table 5-3 Genetic algorithm parameter settings

| Parameters | Generations | Population size | Gene length |
|---|---|---|---|
| Value | 50 | 20 | 10 |

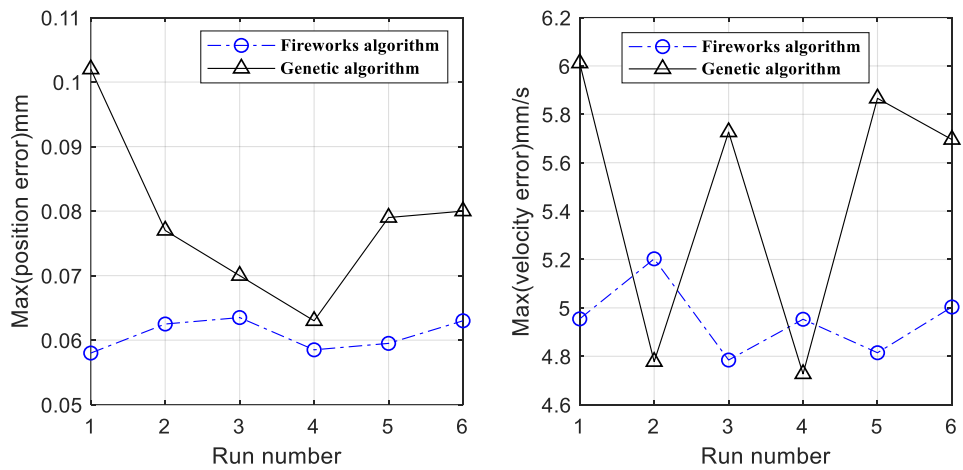

Figure 5-3 Fluctuations in system performance

5.2 Verification of decoupling methods for control parameters

The system control parameters corresponding to each group of motors were optimized three times, and the resulting control parameter values are shown in Table 5-2.

Table 5-2 Servo parameter optimization results

| Serial number | Control parameters Motor Sequence | $K_p$ | $K_{vp}$ | $K_{vi}$ | $K_{fv}$ |
|---|---|---|---|---|---|
| 1st optimization result | 1 | 24.62 | 4.99 | 2.40 | 0.99 |
| | 2 | 25.17 | 3.64 | 0.23 | 0.98 |
| | 3 | 51.58 | 4.37 | 14.25 | 0.95 |
| | 4 | 50.93 | 3.27 | 17.38 | 0.91 |
|  | 5 | 30.30 | 4.39 | 2.88 | 0.98 |
|  | 6 | 69.54 | 4.42 | 1.47 | 0.85 |
| 2nd optimization result | 1 | 47.32 | 0.34 | 20 | 0.50 |
|  | 2 | 39.49 | 0.56 | 19.99 | 0.73 |
|  | 3 | 54.13 | 0.47 | 15.40 | 0.95 |
|  | 4 | 27.12 | 0.51 | 17.53 | 0.95 |
|  | 5 | 5.75 | 0.55 | 20 | 0.95 |
|  | 6 | 5.03 | 0.58 | 11.19 | 0.91 |
| 3rd optimization result | 1 | 87.12 | 0.47 | 19.99 | 0.71 |
|  | 2 | 38.41 | 0.54 | 8.52 | 0.99 |
|  | 3 | 108.77 | 0.48 | 19.99 | 0.78 |
|  | 4 | 113.64 | 0.53 | 7.40 | 0.66 |
|  | 5 | 17.91 | 0.69 | 16.35 | 0.92 |
|  | 6 | 9.29 | 0.79 | 13.63 | 0.84 |

Based on the optimized control parameters, the changes in system performance corresponding to the three control parameter optimizations were analyzed. The results are given in Figure 5-4.

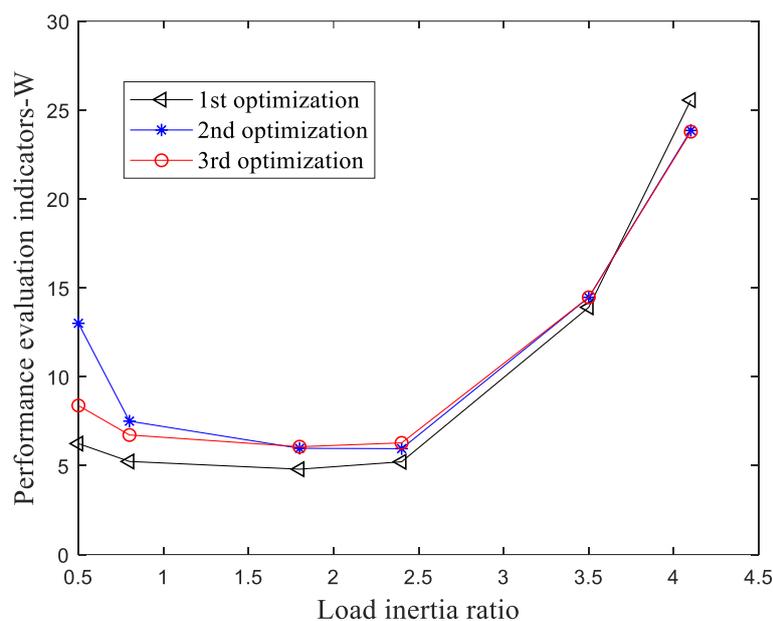

Figure 5-4 Effect of inertia ratio on system performance based on control parameter optimization

The figure above shows the performance changes of the feed system servo control parameters after the optimization. The horizontal coordinate is the load inertia ratio and the vertical coordinate represents the system comprehensive evaluation index W. Line shapes correspond to different optimization results. By comparing these three optimization results, it is evident that the differences in system performance in terms of values are not significant. The overall change pattern is as follows: when the feed system inertia ratio is too large, the system performance sharply deteriorates. On the other hand. when the feed system inertia ratio is too small, the system performance declines. For this reason, it is necessary to limit the inertia ratio range to ensure the normal operation of the system. Therefore, by optimizing the control parameters, the interference of the control parameters on the electromechanical coupling can be eliminated, making it possible to obtain a uniform law of the effect of electromechanical coupling characteristics on the system performance.

The results show that it is indeed possible to optimize the control parameters to achieve the decoupling of the control parameters and eliminate the influence of the control parameters on the system performance. The influence of the coupling relationship between electromechanical subsystems on the dynamic performance of the system was especially highlighted. The findings allow us to remove the influence of the control parameters and only consider the effect of the electromechanical subsystem on the dynamic performance of the feed system. Further, it also can guide the dynamic design of machine tool feed system.

# 6  Experimental validation

In this study, an experimental platform for the feeding system was constructed to verify the impact of decoupling the control parameters. The platform consists of a motor, controller, and mechanical components, as depicted in Figure 6-1. The NI PCI-6321 board is used to transmit signals between the experimental platform and the computer. By eliminating the ball screw link, the platform allows the motor's output axis to directly drive the load movement. Additionally, the motor can be easily replaced, facilitating the investigation of the coupling between subsystems and the verification of the control parameter decoupling method.

To facilitate the comparison and validation of simulation and experimental results, three motors with parameters similar to those used in the simulation process were employed in this experiment. The relevant parameters of the motors used in the experiment are presented in Table 6-1.

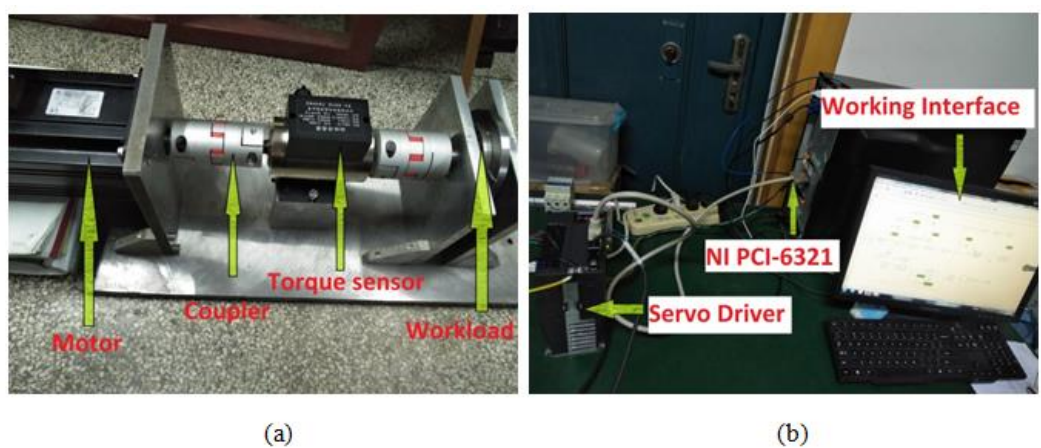

Figure 6-1 Pictures of the experimental platform.

(a) Experimental setup (b) Data acquisition and processing

Table 6-1 Parameters of the motor used in the experimental platform

| Serial number | Motor Model | Maximum Torque T/N m | Rotor inertia $J_m$/kg cm$^2$ | Inertia Ratio |
|---|---|---|---|---|
| 1 | ISMH3-29C15CD | 37.2 | 55 | 0.8 |
| 2 | ISMH3-18C15CD | 28.75 | 25.5 | 1.9 |
| 3 | ISMH3-85B15CD | 13.5 | 13 | 3.7 |

The load applied in this experimental setup was determined to be 48 kg·cm$^2$, which closely aligns with the simulated value. The system's inertia ratio is modified by substituting the motor. The employed control algorithm remains consistent for both the experiment and simulation, and identical control parameters are utilized for a given inertia ratio in both cases. Multiple sets of experiments are conducted by modifying the motion process, providing additional confirmation of the accuracy of the control parameter decoupling method described in this paper.

6.1 Experiments at different accelerations

Based on the motion process outlined in section 2.3, the maximum feed rate is set at 100mm/s. To conduct experiments, acceleration values of 0.5 m/s$^2$, 1 m/s$^2$, 2 m/s$^2$, and 5m/s$^2$ were selected. The outcomes of these experiments are illustrated in Figure 6-2 to Figure 6-5.

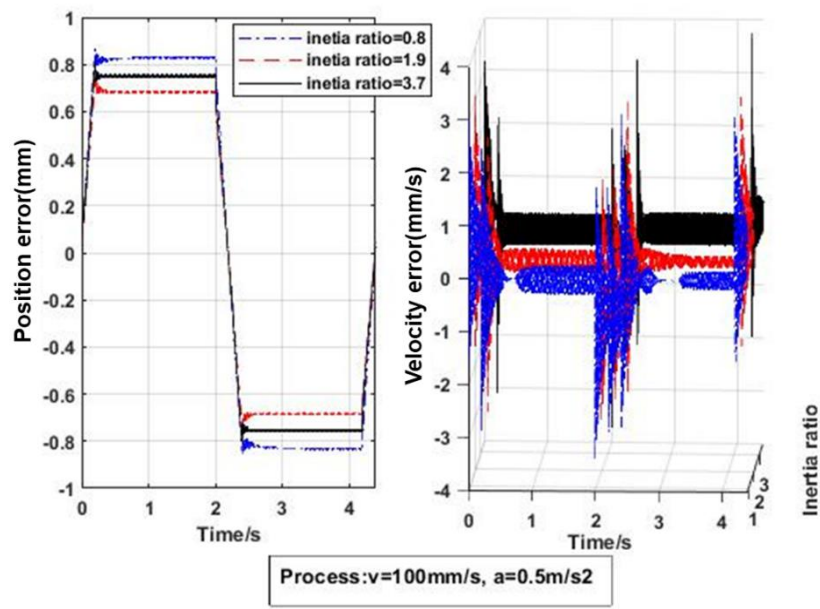

Figure 6-2 Position error and velocity error curves, a=0.5m/s$^2$

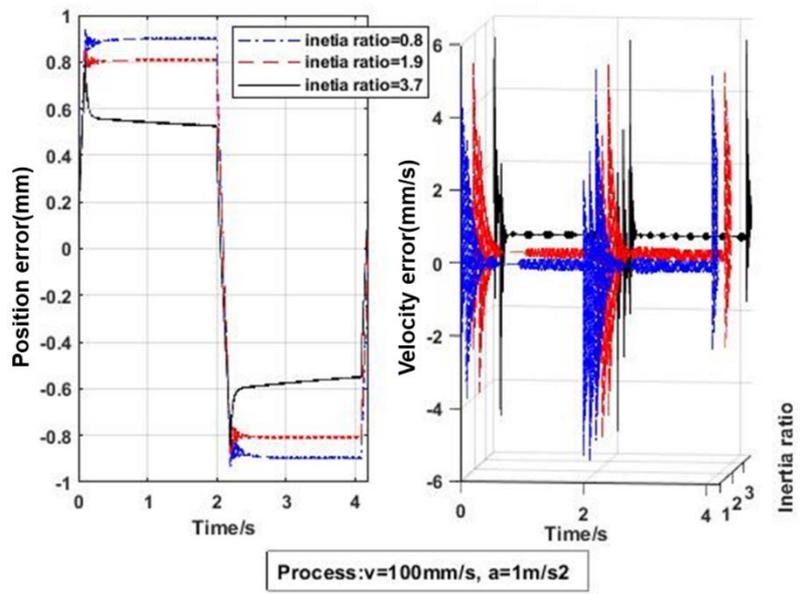

Figure 6-3 Position error and velocity error curves, a=1m/s$^2$

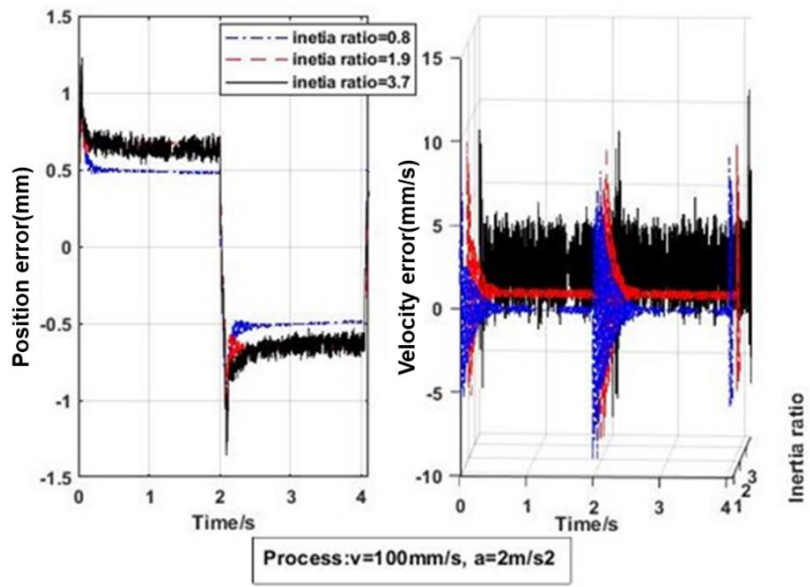

Figure 6-4 Position error and velocity error curves, a=2m/s$^2$

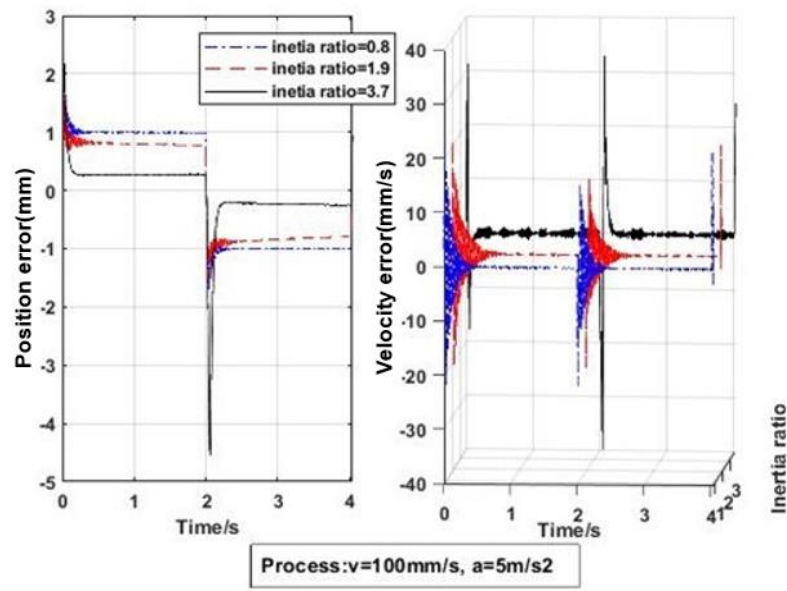

Figure 6-5 Position error and velocity error curves, a=5m/s$^2$

To enable comparison with the simulation results, the peak position error, peak velocity error, and velocity fluctuation of the four experimental processes were extracted based on the performance indices defined in Section 3. These values are presented in Table 6-2. Subsequently, the performance indices for the aforementioned four experimental processes were calculated and compared to the simulation results, as illustrated in Figure 6-6.

Table 6-2 Dynamic performance of the system under different accelerations and inertia ratios

| Acceleration (m/s$^2$) | Inertia ratio | Position error (mm) | Speed error (mm/s) | Speed fluctuation |
|---|---|---|---|---|
| 0.5 | 0.8 | 0.8602 | 3.4166 | 0.5246 |
|  | 1.9 | 0.7460 | 3.1110 | 0.4552 |
|  | 3.7 | 0.8109 | 3.7626 | 0.4286 |
| 1 | 0.8 | 0.9370 | 5.4033 | 0.6121 |
|  | 1.9 | 0.8785 | 5.2174 | 0.5879 |
|  | 3.7 | 0.8548 | 5.4105 | 0.4258 |
| 2 | 0.8 | 1.2320 | 9.1611 | 0.8416 |
|  | 1.9 | 1.1390 | 8.9667 | 0.8471 |
|  | 3.7 | 1.3570 | 10.718 | 1.9117 |
| 5 | 0.8 | 1.8060 | 22.2977 | 2.5519 |
|  | 1.9 | 1.633 | 20.9221 | 2.4685 |
|  | 3.7 | 4.5325 | 39.7184 | 2.3317 |

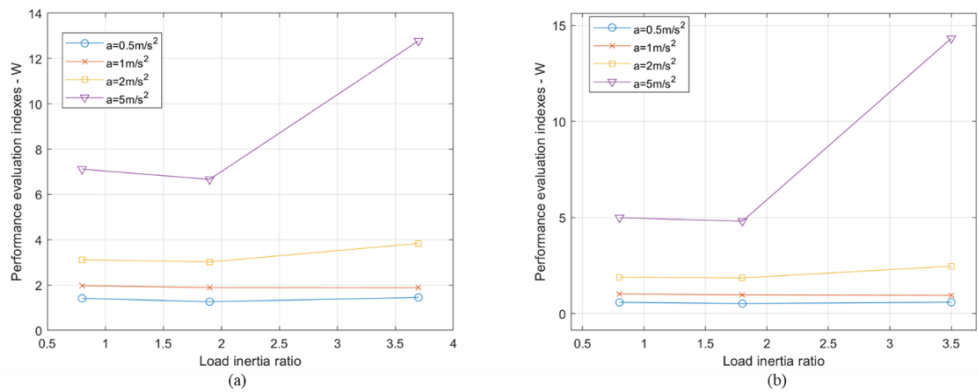

Figure 6-6 The effect of inertia ratio on the performance of the system under different accelerations

after decoupling the control parameters. (a) Experimental results; (b) Simulation results

Based on the experimental results presented in Figure 6-6, it is observed that decoupling the control parameters minimizes the impact of the inertia ratio on system performance under low acceleration conditions. Nevertheless, when the acceleration exceeds a certain threshold, it becomes vital to confine the inertia ratio within a specific range to ensure optimal dynamic performance of the system.

6.2 Experiments at different speeds

Based on the motion process described in Section 2.3, the acceleration rate is set at $1m/s^2$. To evaluate performance, four sets of experiments were conducted with varying maximum feed speeds ranging from 50mm/s to 400mm/s. The outcomes of these experiments are portrayed in Figure 6-7 to Figure 6-10.

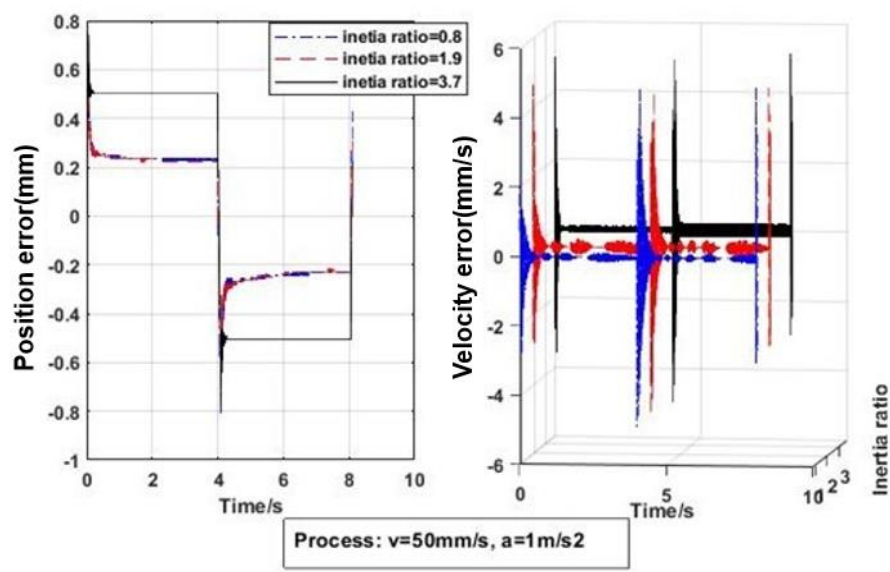

Figure 6-7 Position error and velocity error curves v=50mm/s

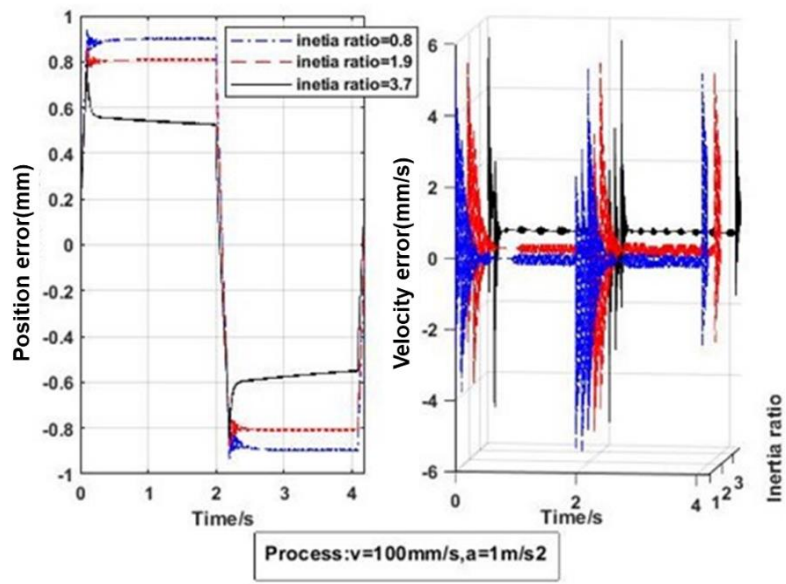

Figure 6-8 Position error and velocity error curves v=100mm/s

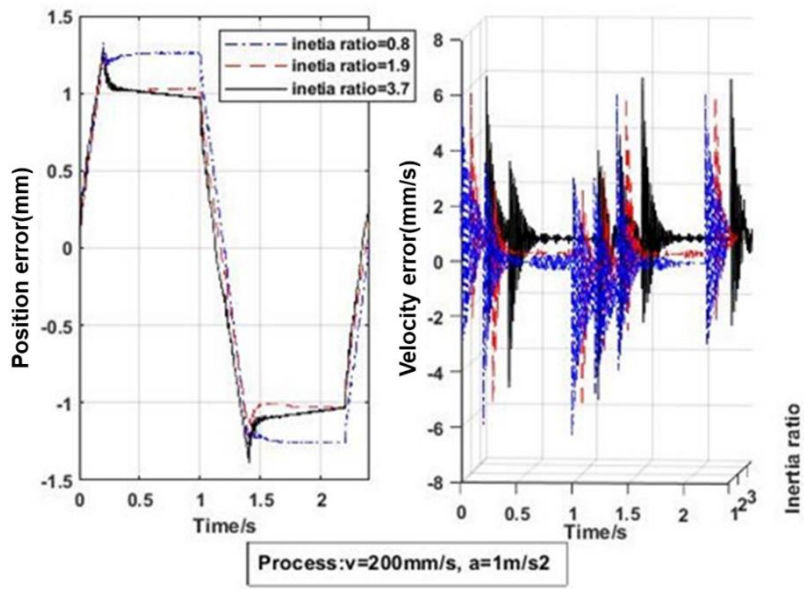

Figure 6-9 Position error and velocity error curves v=200mm/s

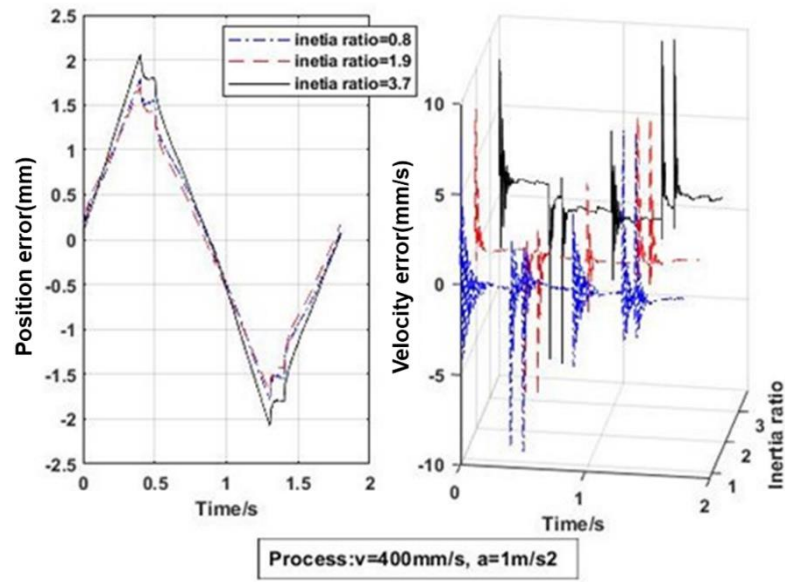

Figure 6-10 Position error and velocity error curves v=400mm/s

To enable comparison with the simulation results, the peak position error, peak velocity error, and velocity fluctuation of the four experimental processes were extracted based on the performance indices defined in Section 3. These values are presented in Table 6-3. Subsequently, the performance indices for the aforementioned four experimental processes were calculated and compared to the simulation results, as illustrated in Figure 6-11.

Table 6-3 Dynamic performance of the system under different speeds and inertia ratios

| Speed (mm/s) | Inertia ratio | Position error (mm) | Speed error (mm/s) | Speed fluctuation |
|---|---|---|---|---|
| 50 | 0.8 | 0.8080 | 4.9404 | 0.2855 |
| | 1.9 | 0.6955 | 4.8327 | 0.2848 |
| | 3.7 | 0.7775 | 5.1419 | 0.2515 |
| 100 | 0.8 | 0.9370 | 5.4033 | 0.6121 |
| | 1.9 | 0.8785 | 5.2174 | 0.5879 |
| | 3.7 | 0.8548 | 5.4105 | 0.4258 |
| 200 | 0.8 | 1.3220 | 6.3113 | 1.1318 |
| | 1.9 | 1.2330 | 5.7150 | 0.8243 |
| | 3.7 | 1.3825 | 5.5890 | 1.0249 |
| 400 | 0.8 | 1.7895 | 9.0494 | 1.3850 |
| | 1.9 | 1.7050 | 7.9003 | 1.0922 |
| | 3.7 | 2.0635 | 9.1434 | 1.1567 |

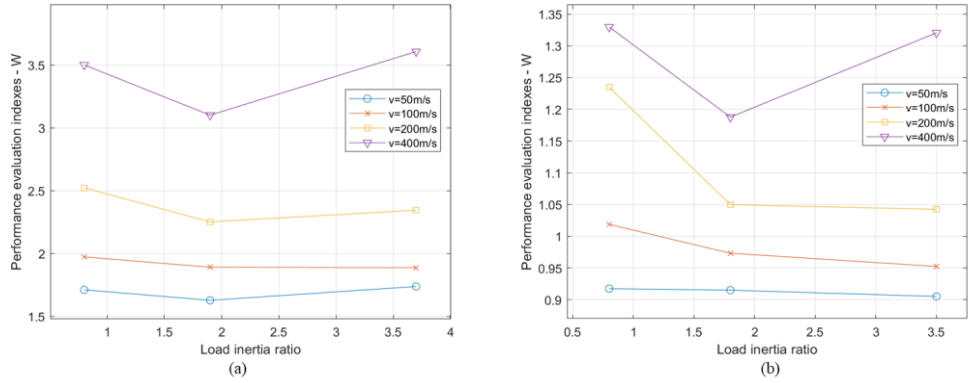

Figure 6-11 The effect of inertia ratio on the performance of the system under different speeds after

decoupling the control parameters. (a) Experimental results; (b) Simulation results

From the experimental results described in Figure 6-11, it can be seen that after decoupling the control parameters, the inertia ratio has little influence on the system performance when the feed rate is small. However, when the feed rate is too large, it is necessary to limit the inertia ratio to a certain range, so as to ensure that the system has good dynamic performance.

The results depicted in Figures 6-6 and 6-11 demonstrate that the feed system model described in this paper, which fails to consider factors like friction, leads to a slightly inferior actual system dynamic performance compared to the experimental results. Interestingly, the change in system dynamic performance, simulation results, and experimental results remain consistent across various motion processes. This finding demonstrates that the control parameter decoupling algorithm utilized in the simulation model outlined in chapters 2 to chapter 5 has a similar effect in the actual experiment. Moreover, this verification affirms the efficacy of the control parameter decoupling method in actual experimental research scenarios, demonstrating its considerable impact in decoupling under various motion processes. Furthermore, this enables the researchers to investigate the coupling between the subsystems of the feed system more comprehensively by eliminating the influence of control parameters. Ultimately, this establishment serves as a solid foundation for uncovering the coupling mechanism among the subsystems of the feeding system and exploring a quantitatively comprehensive optimization design scheme for the feeding system.

# 7 Conclusion

In this paper, the authors studied whether control parameters affect the action law of electromechanical coupling characteristics on system performance. In the next step, it was attempted to eliminate their influence on the electromechanical coupling relationship. The following conclusions were made based on the results:

(1) The system performance trends corresponding to different inertia ratios change with the servo control parameters of the feed system. In other words, the control parameters affect how electromechanical coupling characteristics affect the system performance. Hence, the effects of control parameters on the electromechanical coupling characteristics must be eliminated to study the coupling characteristics between the motor and the mechanical structure.

(2) Elimination of the control subsystem influence on the electromechanical coupling

relationship through optimizing the servo control parameters was proposed. This is done to achieve the decoupling of the control system and highlight the influence of the electromechanical subsystem on the dynamic performance of the overall system. The results have shown that the optimization of the servo control parameters mitigates the interference of the control parameters on the electromechanical coupling characteristics. The system performance variation law remains unchanged, confirming the method's feasibility.

This paper provides a reliable method to remove the influence of the control parameters. It should be noted that only consider the effect of the electromechanical subsystem on the dynamic performance of the feed system was considered. Furthermore, a possibility for further research on the matching characteristics of the feed system subsystems was given. Additionally, it facilitates further study of the coupling mechanism of subsystems, revealing the underlying mechanism behind the dynamic design of feed systems.

# References

1. Erkorkmaz K, Altintas Y. High speed CNC system design. Part II: modeling and identification of feed drives. Int J Mach Tools Manuf. 2001 Aug 1;41(10):1487–509.

2. Wang X, Zhang D, Zhang Z. A review of dynamics design methods for high-speed and high-precision CNC machine tool feed systems [Internet]. arXiv; 2023 [cited 2023 Jul 10]. Available from: http://arxiv.org/abs/2307.03440

3. M. S. Dorney, B. L. Wilson, J. R. Shadley. Unstable Self-Excitation of Torsional Vibration in AC Induction Motor Driven Rotational Systems. J Vib Acoust. 1992 Apr 1;114:226–31.

4. Ferretti G, Magnani G, Rocco P. Experimental analysis of the disturbances affecting contact force in industrial robots. Proc Int Conf Robot Autom Robot Autom 1997 Proc 1997 IEEE Int Conf On. 1997 Jan 1;3:2184.

5. Szolc T, Konowrocki R, Michajłow M, Pręgowska A. An investigation of the dynamic electromechanical coupling effects in machine drive systems driven by asynchronous motors. Mech Syst Signal Process. 2014 Dec 20;49(1):118–34.

6. Lu B( 1 2 ), Zhao W( 1 2 ), Zhang J( 1 2 ), Yang X( 1 2 ), Wang L( 1 2 ), Zhang H( 1 2 ), et al. Electromechanical coupling in the feed system with high speed and high acceleration. Jixie Gongcheng XuebaoJournal Mech Eng. 2013 Mar 20;49(6):2–11.

7. Yang X, Lu D, Zhang J, Zhao W. Dynamic electromechanical coupling resulting from the air-gap fluctuation of the linear motor in machine tools. Int J Mach Tools Manuf. 2015 Jul 1;94:100–8.

8. Yang X( 1 2 ), Lu D( 1 2 ), Ma C( 1 2 ), Zhang J( 1 2 ), Zhao W( 1 2 ). Analysis on the multi-

dimensional spectrum of the thrust force for the linear motor feed drive system in machine tools. Mech Syst Signal Process. 2017 Jan 1;82:68–79.

9. Yang X, Liu H, Lu D, Zhao W. Investigation of the dynamic electromechanical coupling due to the thrust harmonics in the linear motor feed system. Mech Syst Signal Process. 2018 Oct 1;111:492–508.

10. Yang X, Lu D, Zhao W. Decoupling and effects of the mechanical vibration on the dynamic precision for the direct-driven machine tool. Int J Adv Manuf Technol. 2018 Apr;95(9–12):3243–58.

11. Yang X, Song B, Xuan J. Effects of the mechanical vibrations on the thrust force characteristics for the PMLM driven motion system. Mech Syst Signal Process. 2022 Aug 1;175:109110.

12. Kim MS, Chung SC. Integrated design methodology of ball-screw driven servomechanisms with discrete controllers. Part I: Modelling and performance analysis. Mechatronics. 2006 Oct;16(8):491–502.

13. Yang Xiaojun, Wanhua Zhao, Kun Cao, Hui Liu, Jun Zhang, Huijie Zhang. Research on Mechatronic Coupling Facts of Linear Motor Feed Drive System Based on Spectrum Characteristics. In: Volume 4: Dynamics, Control and Uncertainty, Parts A and B. American Society of Mechanical Engineers; 2012.

14. Zhang T, Zhang D, Zhang Z, Muhammad M. Investigation on the load-inertia ratio of machine tools working in high speed and high acceleration processes. Mech Mach Theory. 2021 Jan 1;155.

15. Feng B( 1 ), Mei X( 1 ), Mu E( 1 ), Huang X( 1 ), Zhang D( 2 ). Investigation of the controller parameter optimisation for a servomechanism. Proc Inst Mech Eng Part B J Eng Manuf. 2015 Jan 1;229:98–110.

16. Li Wu, Xiu Xia Yu, Yi Chen Wang. Selection and Analysis of Servomotor for Three-Axis Transmission System in CNC Machine Tool. Adv Mater Res. 2013 Sep 18;1148–53.

17. Bin Bin Qian, Jun Sun, Xian Jun Qin, Yuan Huang. Modeling and Simulation of Kollmorgen Motor Movement Control System Based on SimScape. Appl Mech Mater. 2015 Jan 1;872–5.

18. Andolfatto L, Lavernhe S, Mayer J r. r. Evaluation of servo, geometric and dynamic error sources on five-axis high-speed machine tool. Int J Mach Tools Manuf. 2011 Jan 1;51(10):787–96.

19. Cheng J, Ji J, Guo Y nan, Ji J. Dynamic Multiobjective Software Project Scheduling Optimization Method Based on Firework Algorithm. Math Probl Eng. 2019 7/1/2019;1–13.

20. Yang W, Ke L. An improved fireworks algorithm for the capacitated vehicle routing problem. Front Comput Sci. 2019;13:552–64.